\documentclass[a4paper,11pt]{article}
\pdfoutput=1 

\usepackage{jheppub} 

\usepackage[T1]{fontenc} 

\title{On the Lagrangian and fermion mass of the 
  unified $SU(2) \otimes SU(4)$ gauge field theory}

\author[a]{Eckart Marsch,}
\author[b,c]{and Yasuhito Narita}

\affiliation[a]{Johann-Fleck-Stra{\ss}e 18, Kiel 24106, Germany}
\affiliation[b]{Institut f\"ur Theoretische Physik, 
  Technische Universit\"at Braunschweig, 
  Mendelssohnstra{\ss}e 38, 38106 Braunschweig, Germany}
\affiliation[c]{Max Planck Institute for Solar System Research, 
  Justus-von-Liebig-Weg 3, 37077 G\"ottingen, Germany}

\emailAdd{eckart.marsch@web.de}
\emailAdd{y.narita@tu-braunschweig.de}

\abstract{In the Standard Model of elementary particles the fermions are assumed to be intrinsically massless. Here we propose a new theoretical idea of fermion mass generation (other than by the Higgs mechanism) through the coupling with the vector gauge fields of the unified $SU(2) \otimes SU(4)$ gauge symmetry, especially with the $Z$ boson of the weak interaction that affects all elementary fermions. The resulting small masses are suggested to be proportional to the self-energy of the $Z$ field as described by a Yukawa potential. Thereby the electrically neutral neutrino just gets a tiny mass through its $Z$-field coupling. In contrast, the electrically charged electron and quarks can become more massive by the inertia induced through the Coulomb energy of the electrostatic fields surrounding them in their rest frames.}

\begin{document} 
\maketitle
\flushbottom


\section{Introduction}
\label{sec1}

According to the common wisdom of the Standard Model (SM) \cite{schwartz2014,yangmills1954} of elementary particle physics, the fermions are intrinsically massless, but they gain their masses via phase transition from the vacuum of the Higgs field \cite{higgs1964,weinberg1967}. However, this notion introduces many free parameters (the Yukawa coupling constants) that are to be determined through measurements. These have been made at the LHC only for some members of the second and third family of heavy leptons and quarks, yet not for the important first family of fermions, of which the stable and long-lived hadrons form according to the gluon forces of quantum chromodynamics (QCD).

Here we just consider the first fermion family of the SM and propose a new idea of the fermion mass generation. The key assumption is that their masses may be equal to the relevant gauge-field energy in the rest frames of these charged fermions carrying electroweak or strong charges. Their masses are suggested to originate from jointly breaking the chiral $SU(2)$ symmetry combined with the hadronic isospin $SU(4)$ symmetry, as described in the recent model by Marsch and Narita \cite{mana2024}, following early ideas of Pati and Salam \cite{patisalam1973,patisalam1974} and their own work \cite{mana2016,mana2022}. Unlike in the SM, in their model both symmetries are considered as being unified to yield the $SU(2)\otimes SU(4)$ symmetry, which then is broken by the same procedures that are applied successfully in the electroweak sector of the SM. 

The outline of the paper is as follows. We briefly discuss the extended Dirac equation \cite{mana2024,dirac1928} and its Lagrangian including the Higgs, gauge-field and fermion sectors. Especially, the covariant derivative is discussed and
the various gauge-field interactions are described. Also the different charge operators (weak and strong) are presented. Then the CPT theorem is derived for the extended Dirac equation including the gauge field terms. The remainder of the paper addresses the idea of mass generation from gauge field energy in the fermion rest frame. Finally we present the conclusions.


\section{The extended Dirac equation including the unified gauge fields and a mass term}
\label{sec:2}

First we quote here the extended Dirac equation after Marsch and Narita \cite{mana2024}, which includes the unified gauge fields but also a single mass $m$ in terms of a diagonal mass matrix $\texttt{M}=m {\sf{1}_8}$ as follows
\begin{equation}
	\label{eq:17} 
	(\gamma^\mu \otimes \mathrm{i} \texttt{D}_\mu) \Psi(x) = (\gamma^\mu \otimes (\mathrm{i} \partial_\mu {\sf{1}_8}  + \texttt{S}_\mu(x)))\Psi(x) = ({\sf{1}_4} \otimes \texttt{M})\Psi(x).
\end{equation}
In the main part of this paper we will propose and derive a new less simple mass matrix stemming from energy of the gauge field $\texttt{S}_\mu$ in the fermion rest frame. Here $\gamma^\mu$ is the standard $4 \times 4$ Dirac matrix four-vector, and $\Psi(x)$ is an extended 32-component Dirac spinor. The spacetime coordinate is $x=x^\mu=(t,\mathbf{x})$. Units are such that $c=\hbar=1$. 

The gauge field of the combined $SU(2) \otimes SU(4)$ symmetry is denoted by an $8 \times 8$ hermitian matrix-field $\texttt{S}_\mu$, which involves the two coupling constants $g_2$ and $g_4$ for the two independent gauge fields. After chiral symmetry breaking, we can concisely write the symmetry-related part of the above covariant derivative as
\begin{equation}
	\label{eq:18}
	\texttt{S}_\mu(x) = \texttt{Q} A_\mu(x) + \texttt{R} Z_\mu(x) + g_2 \texttt{W}_\mu(x) + g_4 \texttt{V}_\mu(x)
	+ g_4 \texttt{C}_\mu(x).
\end{equation}
The diagonal matrix charge-operators $\texttt{Q}$ and $\texttt{R}$ are presented below in equations 
(\ref{eq:6}) and (\ref{eq:11}) in terms of numbers. The matrix fields $\texttt{V}_\mu$ (``leptoquark" gauge field for lepton-quark exchange) and $\texttt{W}_\mu$ (for flavour transitions associated with electric charge exchange) are associated with the massive $V$ and $W$ bosons, and the matrix field $\texttt{C}_\mu$ is related to the quarks only as in QCD. $A_\mu$ is the electromagnetic vector potential (massless photon) coupling to electrically charged fermions only, and  $Z_\mu$ is the gauge field associated with the heavy $Z$ boson. All these terms were derived and given explicitly in \cite{mana2024}. The matrix fields are hermitian.


\section{The Lagrangian of the unification model for the SU(2) and SU(4) symmetries}
\label{sec:3}

The Lagrangian of the unified weak-strong interactions consists like in the SM of three parts related to the Higgs, the unified Fermion and the massless and massive heavy gauge bosons.
\begin{equation}
	\label{eq:441}
	\mathcal{L} = \mathcal{L}_\mathrm{h} + \mathcal{L}_\mathrm{f} + \mathcal{L}_\mathrm{g}. 
\end{equation}
The scalar Higgs field $\Phi(x)$ has 8 components and its Lagrangian reads
\begin{equation}
	\label{eq:442}
	\mathcal{L}_\mathrm{h} = ( \texttt{D}^\mu \Phi )^\dagger 
	(\texttt{D}_\mu \Phi ).
\end{equation}
Inserting here just the Higgs vacuum ($v$) field $\Phi_v=v(0,0,0,1,0,0,0,0)$ delivers after {\cite{mana2024} the masses of the $V$, $W$ and $Z$ bosons. The spinorial unified Fermion field $\Psi(x)$ has 32 components and its Lagrangian reads
\begin{equation}
	\label{eq:443}
	\mathcal{L}_\mathrm{f} = \overline{\Psi} (\mathrm{i} \texttt{D}^\mu - \texttt{M})\Psi. 
\end{equation}
Here the conjugate fermion spinor reads as usually $\overline{\Psi}=( (\gamma_0 \otimes {\sf{1}_8}) \Psi )^\dagger$. In the SM there is an additional linear Yukawa coupling of the fermion with the Higgs field \cite{weinberg1967,schwartz2014}, which we shall not consider here, as the main purpose of this paper is to suggest another mechanism owing to the gauge fields involved. In the remainder of this work we derive a new idea for the origin of the mass matrix $\texttt{M}$. Finally, the Lagrangian of the weak and strong gauge-field matrix tensor $F^{\mu\nu}(x)$ reads
\begin{equation}
	\label{eq:444}
	\mathcal{L}_\mathrm{g} = - \frac{1}{4\pi} \texttt{F}^{\mu\nu} \texttt{F}_{\mu\nu}. 
\end{equation}
We recall that the field tensor is here an $8 \otimes 8$ matrix given by the standard expression of Yang-Mills theory \cite{schwartz2014,yangmills1954} 
\begin{equation}
	\label{eq:445}
	\texttt{F}_{\mu\nu} = \partial_\mu \texttt{S}_\nu - \partial_\nu \texttt{S}_\mu - \mathrm{i} [\texttt{S}_\mu, \texttt{S}_\nu]. 
\end{equation}
Concerning the commutator involved, we note that the first four terms in (\ref{eq:24}) further below commute with each other, yet they all do not commute with $\texttt{I}_\mu$, which does also not commute with itself.


\section{The gauge-field interactions matrix}
\label{sec:4}

Here we present again the above interaction term $\texttt{S}_\mu$ after the equation (\ref{eq:18}) as it appears in the covariant derivative $D_\mu$ in equation (\ref{eq:17}) in terms of an $8 \otimes 8$ matrix. For that purpose we reorder the last three terms after Marsch and Narita \cite{mana2024}  (given explicitly in their equations (65)-(72)), and we also define the hadronic charge operators which can be expressed by diagonal matrices as follows 
\begin{equation}
	\label{eq:22}
	\texttt{Q}_2 = \frac{g_4}{2} \,\mathrm{diag}\,[1,-1,0,0,1,-1,0,0].
\end{equation}
\begin{equation}
	\label{eq:23}
	\texttt{Q}_3 = \frac{g_4}{2} \,\mathrm{diag}\,[1,1,-2,0,1,1,-2,0].
\end{equation}
Obviously, the leptons (entries 4 and 8) carry no strong hadronic charge and thus are decoupled from the strong interaction. Here $\texttt{Q}_2$ is the charge operator associated with the $SU(2)$ subgroup of $SU(3)$, and $\texttt{Q}_3$ with the $SU(3)$ subgroup of $SU(4)$. The corresponding gluon fields, $G^2_\mu$ and $G^3_\mu$, are akin to the electric field $E_\mu$ of the massless photon. All the remaining fields correspond to rotations in the complex Hilbert space $C^4$ related to the original $SU(4)$ group, and therefore they link the eight fermions between each other. Then we obtain the interaction part of the covariant derivative as follows
\begin{equation}
	\label{eq:24}
	\texttt{S}_\mu = \texttt{Q} A_\mu + \texttt{R} Z_\mu + \texttt{Q}_2 G^2_\mu + \texttt{Q}_3 \frac{1}{\sqrt{3}} G^3_\mu + \texttt{I}_\mu.
\end{equation} 
The first four terms have diagonal charge matrices, whereas the last terms is a complex $8 \otimes 8$ matrix, containing all the weak and strong exchange interactions combined. Note that $\texttt{I}_\mu$ is a hermitian trace-less gauge-field matrix, which mixes the quarks and leptons among each other and is most important for hadronic binding and quarks confinement. To ease the notation we omit here in the matrix the subscript $\mu$ indicating a relativistic four-vector. Then we obtain for the gauge-field matrix 
$\texttt{I}_\mu =\frac{g_4}{2} \times$
\begin{equation}
	\label{eq:25}
	\left(
	\begin{array}{cccccccc}
		0 & G^-_1 & G^-_2 & V^-_1 & \frac{g_2}{g_4} W^- & 0 & 0 & 0 \\
		G^+_1 & 0 & G^-_3 & V^-_2 & 0 & \frac{g_2}{g_4} W^- & 0 & 0 \\
		G^+_2 & G^+_3 & 0 & V^-_3 & 0 & 0 & \frac{g_2}{g_4} W^- & 0 \\
		V^+_1 & V^+_2 & V^+_3 & 0 & 0 & 0 & 0 & \frac{g_2}{g_4} W^- \\
		\frac{g_2}{g_4} W^+ & 0 & 0 & 0 & 0 & G^-_1 & G^-_2 & V^-_1 \\
		0 & \frac{g_2}{g_4} W^+ & 0 & 0 & G^+_1 & 0 & G^-_3 & V^-_2 \\
		0 & 0 & \frac{g_2}{g_4} W^+ & 0 & G^+_2 & G^+_3 & 0 & V^-_3 \\
		0 & 0 & 0 & \frac{g_2}{g_4} W^+ & V^+_1 & V^+_2 & V^+_3 & 0 \\
	\end{array}
	\right). 
\end{equation}
It is obvious by inspection that the matrix is hermitian, $\texttt{I}^\dagger_\nu= \texttt{I}_\mu$, according to the definitions of the gauge fields in \cite{mana2024}. Whereas the six gluon gauge fields $G^\pm_{1,2,3}$ are massless, the other boson fields $(V_{1,2,3}^\pm)_\mu$ and $W^\pm_\mu$ are rather heavy owing to the Higgs mechanism \cite{schwartz2014,mana2024}. In the SM the six heavy gauge bosons $V^\pm_n$ (``leptoquarks'') are not considered but omitted in the SM Lagrangian. Then inspection of (\ref{eq:25}) reveals that the QCD subgroup $SU(3)$ emerges. The respective four block matrices of the weak and strong interactions can clearly be identified in (\ref{eq:25}). They appear as well separated blocks, since the full $8 \times 8$ matrix was constructed as a Kronecker product.


\section{The fermion charges related to the photon and $Z$ boson}
\label{sec:5}

In their recent paper Marsch and Narita \cite{mana2024} developed a unified gauge theory for elementary fermions based on the Kronecker product of the gauge groups $SU(2) \otimes SU(4)$. When this symmetry is broken by application of the Higgs mechanism  \cite{higgs1964}, the $Z$-boson becomes massive and couples to all eight elementary fermions which carry corresponding charges described by the diagonal operator named $\texttt{R}$, which is the analogue to the electromagnetic charge operator 
\begin{equation}
	\label{eq:6}
	\texttt{Q} = e\,\mathrm{diag}\,[+\frac{2}{3},+\frac{2}{3},+\frac{2}{3},0,-\frac{1}{3}, -\frac{1}{3}
	,-\frac{1}{3},-1 ].
\end{equation}  
Here the electromagnetic charge unit $e$ is defined by the two involved coupling constants, $g_2$ for the weak and $g_4$ for the strong interaction, which are associated before symmetry breaking with the original gauge groups $SU(2)$, respectively, $SU(4)$. One finds that
\begin{equation}
	\label{eq:7}
	e = \frac{g_2g_4}{\sqrt{g_4^2 + \frac{2}{3} g_2^2}}.
\end{equation}
For the charges related with the electroweak field $Z_\mu$ one obtains
\begin{equation}
	\label{eq:8}
	\texttt{R} = \frac{e}{2}\,\mathrm{diag}\,[q_-,q_-,q_-,l_-, q_+,q_+,q_+,l_+].
\end{equation}
The $q_\pm$ correspond to the three up and down quarks, and $l_\pm$ to the neutrino and electron. Their values are
\begin{equation}
	\label{eq:9}
	l_\pm = \pm \sqrt{\frac{2}{3}} \frac{g_2}{g_4} -\sqrt{\frac{3}{2}} \frac{g_4}{g_2}.
\end{equation}
Similarly, one obtains
\begin{equation}
	\label{eq:10}
	q_\pm = \pm \sqrt{\frac{2}{3}} \frac{g_2}{g_4} + \frac{1}{3} \sqrt{\frac{3}{2}} \frac{g_4}{g_2}.
\end{equation}
Therefore, the trace of the operator $\texttt{R}$ sums up to zero, as it should by charge conservation.
It turns out in \cite{mana2024} that $g_2 \approx 2g_4$. Inserting that values in (\ref{eq:9}) and 
(\ref{eq:10}), one obtains from (\ref{eq:8})
\begin{equation}
	\label{eq:11}
	\texttt{R} = e\,\mathrm{diag}\,[-0.7,-0.7,-0.7,-1.1, 0.9,0.9,0.9,0.5].
\end{equation}
These values for the weak charges are used in what follows.


\section{The canonical electrodynamic momentum and the classical electron radius}
\label{sec:6}

In the above discussion the mass matrix $\texttt{M}$ remained
undetermined. In the simplest case we may just assume all fermion components have the same mass $m$, however this is not in compliance with the empirical knowledge of the quark and lepton masses. In the SM following Weinberg's suggestion \cite{weinberg1967} the elementary fermions get their mass via the Higgs mechanism and are determined by the Higgs vacuum field. Here we shall propose an alternative to this notion by exploiting that the weak and strong gauge fields involved have a finite energy in the rest frame of the respective fermions as induced by their charges.  

According to the special theory of relativity the energy $E$ and momentum $\mathbf{p}$ of a particle with mass $m$ can be combined in the covariant four-vector $p_\mu=(E,\mathbf{p})$, which obeys the relativistic dispersion relation 
\begin{equation}
\label{eq:1}
p^\mu p_\mu = E^2 - \mathbf{p}^2 = m^2, \;\; E(p) = \sqrt{p^2+m^2}.
\end{equation}
Thus the mass is a Lorentz invariant, but what is its physical nature? In classical mechanics and non-relativistic quantum mechanics it is just a given parameter characterizing the particle, like its intrinsic spin $\mathbf{S}$ or electric charge $e$. Yet a charged particle while being in motion carries its electric charge with it, and thus also draws the associated electromagnetic field surrounding it and the related energy with it.
 
Let us for simplicity consider now only classical electrodynamics. When the four-momentum is replaced in (\ref{eq:1}) by the canonically conjugated \cite{weinberg2013} momentum $P_\mu=p_\mu -e A_\mu$, we obtain
\begin{equation}
	\label{eq:111}
	P^\mu P_\mu = (E-eA_0)^2 - (\mathbf{p}-e\mathbf{A})^2 = m^2.
\end{equation}
Taking just the Coulomb potential for $eA_0=\frac{e^2}{r_e}$,
and putting $E$ and the momentum term as well to zero, defines the mass in terms of the electrostatic energy.

We shall therefore conclude here that this property gives an elementary charged particle its inertia, and somehow also defines its physical extent in space. The particle size $r_e$ can be estimated by simple dimensional analysis equating its energy at rest with the Coulomb field energy evaluated at that distance, which is named the classical radius of the charged particle. Thus for the electron, which is known to be elementary, one then obtains (in standard electrostatic units, \cite{jackson1975}) the simple relation (with $c$ being inserted again)  
\begin{equation}
\label{eq:2}
m_e c^2 = \frac{e^2}{r_e}.
\end{equation}
This equation gives the radius in units of the electron (reduced) Compton wavelength as follows
\begin{equation}
\label{eq:3}
r_e = \alpha_e \lambda_e, \;\; \lambda_e=\frac{\hbar}{m_e c},
\end{equation}
with the well-known dimensionless atomic fine-structure constant. It is defined as 
$\alpha_e = e^2/{\hbar\,c}$ and found empirically to have the small numerical value of $1/137=0.0073$. Then $r_e = 2.82 \times 10^{-15}$m, with $\lambda_e=3.86 \times 10^{-13}$m. The reduced Compton wavelength 
of the proton is smaller than this by the factor of the particles mass ratio $m_e/m_p=1/1837$ and amounts to $\lambda_p=2.1 \times 10^{-16}$m. However, is this quantity really a physically meaningful length?  

Namely, as the proton is known from the SM and scattering experiments to be a composite of three quarks, to apply also the simple formula (\ref{eq:2}) seems rather questionable. Yet if we still do it, we may define the ``classical proton radius" as in (\ref{eq:3}). But then we obtain the very small value of $r_p=1.53 \times 10^{-18}$m, which is in stark contradiction to the measured value of the proton radius that is in the range $r_p=(0.84-0.87)10^{-15}$m. However, we may perhaps apply a formula similar to (\ref{eq:2}) to the quarks themselves being the constituents of the proton. This is done in the subsequent section.  

\section{The electrostatic and effective quark masses and resulting radii}
\label{sec:7}

How does this electron result compare with that for quarks on the basis of the Coulomb law yielding the ``classical quark radius"? To evaluate this we remind of the fractional electric charge $Q_u=2/3$\,e for the up and $Q_d=-1/3$\,e for the down quark. Using again the basic equation (\ref{eq:2}) we obtain for quarks
\begin{equation}
	\label{eq:5}
	m_q c^2 = \frac{Q_q^2}{r_q}, \;\; r_q = (Q_q/e)^2 \alpha_e \lambda_q
\end{equation}
Thus we get the following radii:  $r_u = 3.24 \times 10^{-3} \lambda_u = 2.9  \times 10^{-16}$m, for the up quark with a mass of about $m_u=2.2$\,MeV$/c^2$, and 
$r_d = 8.11 \times 10^{-4} \lambda_d = 3.4 \times 10^{-17}$m, for the down quark with a mass of about $m_d=4.7$\,MeV$/c^2$ \cite{pdg2022}. So the heavier down quark is by about an order of magnitude smaller in terms of the ``classical Coulomb size" than the lighter up quark, which has twice the amount of charge as the down quark, and thus 4 times more Coulomb self energy. 
 
However, we have to keep in mind, that according to the Standard Model (SM) of elementary particle physics \cite{schwartz2014} and its ample experimental validation, quarks are the basic corpuscular building blocks of hadrons, i.e. mesons and baryons. Quarks interact via the exchange of massless gluons. In the SM quarks and gluons as well carry colour charges. Gluons mediate the strong interactions among the quarks and enforce their confinement in the hadrons, of which the proton, neutron and pions are the most prominent and important particles. Unlike the electrostatic Coulomb interaction, the strong interaction associated with the non-Abelian gauge group $SU(3)$ cannot by simply described by a static potential. 

According to theory and the results from various scattering experiments the strong fine-structure coupling constant named here $\alpha_s$ is varying (running) as a function of energy and considerably larger than $\alpha_e$. For example, at the energy of 100\,GeV one finds empirically that $\alpha_s=0.12$, which is about 16.4 times $\alpha_e$. Using that high value one may roughly estimate the effective quark radius from the following equations 
\begin{equation}
\label{eq:4}
m_q c^2 = \frac{\alpha_s \hbar\,c}{R_q}, \;\; R_q = \alpha_s \lambda_q.
\end{equation}
Then the related Compton wavelengths are for the up quark 
$\lambda_u=8.95 \times 10^{-14}$m, and for the down quark 
$\lambda_d=4.19 \times 10^{-14}$m, which differ by a factor of about 2 due to the mass ratio of about 2. 
The corresponding radii then are: $R_u = 1.07  \times 10^{-14}$m, and  $R_d = 5  \times 10^{-15}$m, which both are larger by an order of magnitude than the typical nuclear scale of a femtometer. 

What about the electrically chargeless neutrino? Of course, the ``classical neutrino radius" makes no sense, as it does not take part in the electromagnetic interaction. Neither, does it feel the strong force, and thus the above calculation for quarks exploiting $\alpha_s$ does certainly not apply to the neutrino. However, there remains a gauge field that affects all elementary fermions \cite{mana2024}, which is represented by the short-reached massive $Z$-boson with the measured mass of $M_Z=91.2\,$GeV$/c^2$.

\section{The tiny fermion masses and radii as determined by the $Z$-boson Yukawa-field energy}
\label{sec:8}

As has been shown in the paper of Marsch and Narita \cite{mana2024}, all eight elementary fermions interact with the weak $Z$-boson gauge field. We have four different charges as given in (\ref{eq:8}) in abstract form or by numbers in (\ref{eq:11}). It is important to emphasize that they are of the order of $e$, like in the electromagnetic field case (\ref{eq:6}). However, the neutrino's weak charge now is not zero but $-1.1$e, and thus the neutrino is also coupled via the field $Z_\mu$ to all six quarks and the electron. Since the $Z$ boson is very heavy, it has a rather short Compton wavelength given by $\lambda_Z= \hbar/(M_Z c)= \lambda_e\,m_e/M_Z$. The mass ratio is $5.6 \times 10^{-6}$, and therefore one obtains $\lambda_Z=2.16 \times 10^{-18}$m. 

We can now use similar arguments as in the electrostatic case and use again (\ref{eq:111}), where we replace $A_0$ by $Z_0$ of the $Z$-boson and consider a simple Yukawa-type shielded potential $Z_0 = Y(r)$ with a shielding length given by the Compton wavelength $\lambda_Z$. The potential energy of the weak charge $e_j$ then reads
\begin{equation}
\label{eq:12}
e_j Y(r) = e_j\frac{\exp{(-r/\lambda_Z)}}{r}.
\end{equation}
Like in the previous sections, we put the mass $m_j$ of the fermion species $j$ equal to the field's self energy and thus obtain with the normalized radius $x=r/\lambda_Z$ the result
\begin{equation}
\label{eq:13}
m_j = M_Z (\frac{e_j}{e})^2 \alpha_e \frac{\exp{(-x)}}{x}.
\end{equation}
We recall \cite{mana2024} that the Z-boson mass is $M_Z=\frac{1}{2}\upsilon\sqrt{g_2^2 + g_4^2 \frac{3}{2}}$, with the Higgs vacuum expectation value $\upsilon=246$\,GeV. For the weak charges (in units of $e$) squared we get after (\ref{eq:11}) the value 0.49 for the up quarks, 1.21 for the neutrino, 0.81 for the down quarks and 0.25 for the electron. We recall that the atomic fine-structure constant is $\alpha_e =1/137=0.0073$. How do we interpret equation (\ref{eq:13})? 

Given the mass of the neutrino being known from measurements, we can calculate the ``classical fermion radius"  $r_j$ of the species $j$ on the basis of the rather short-ranged weak interaction mediated by the gauge field $Z_\mu$. We recall that one retains the Coulomb field case of (\ref{eq:2}), by letting 
$M_z$ go to zero which corresponds to letting $\lambda_Z$ go to infinity. Let us just consider the case of the neutrino. Then we obtain the mass as
\begin{equation}
\label{eq:14}
m_\nu = 0.8 \frac{\exp{(-x)}}{x} \mathrm{GeV}/c^2.
\end{equation}
For the mass we consider a typical value as estimated by means of the neutrino oscillations, for example
$m_\nu=0.8$\,eV/$c^2$ \cite{katrin2022,kajita2006}. Equation (\ref{eq:14}) reduces to $x=10^9\exp{(-x)}$, the solution of which is about $x=17.8$ and corresponds to a neutrino radius of $r_\nu = 17.8 \lambda_Z = 3.85 \times 10^{-17}$m. This is about a hundredth of the classical Coulomb electron radius. Associated with it is the Thomson cross section, which is $\sigma_\mathrm{T}=8\pi/3 r^2_e$ and amounts to 
$66.5\,\mathrm{fm}^2$. So the cross section for neutrino scattering at the Yukawa potential via the 
$Z$-boson can be estimated to be about $\sigma_\nu = 10^{-4}\sigma_\mathrm{T}$. Similar calculations can be done for the up and down quarks and the electron, which give the same orders of magnitude of their radii. Then we obtain for those particles similarly small masses based solely on the self-energy of weak interaction field $Z_\mu$ of the $Z$-boson. They are composed in the Table~\ref{tab:1}.
 
\begin{table}
\caption{Fermion weak masses ($m_\nu=0.8\,$eV$/c^2$) induced by the Higgs field vacuum energy $\upsilon$}
\label{tab:1}
\begin{tabular}{ccccc}
\hline
        & up quark & neutrino & down quark & electron \\
\hline
Symbol & $\tilde{m}_u$ &  $m_\nu$ & $\tilde{m}_d$ & $\tilde{m}_e$ \\
Formula & $(e_u/e_\nu)^2 m_\nu$ & Eq.\,(\ref{eq:13}) & $(e_d/e_\nu)^2 m_\nu$ & $(e_e/e_\nu)^2 m_\nu $ \\
Value & 0.40  & 1.00  & 0.67  & 0.21  \\
\hline
\end{tabular}
\end{table}

We have used the known mass of the neutrino to infer its radius based on the formula (\ref{eq:13}) that involves the neutrino energy in the weak-$Z$-boson Yukawa potential. This approach is simple but may appear physically naive. Yet we stress that we considered here the dominant, while non-zero weak charge of the neutrino, and not its subtle electromagnetic properties that may emerge in higher order in scattering experiments, although it has no initial electric charge.   

Given our simple considerations, we can see that the minimal masses for all eight elementary fermions are about the same and of the order of $m_\nu$. Assuming they have the same mass, this would imply slightly different radii, as their weak charges are somewhat different. But there is a more interesting important consequence of formula (\ref{eq:13}), namely that the mass is proportional to the $Z$-boson mass, which in turn is proportional to the Higgs vacuum $\upsilon=246$\,GeV according the the Higgs mechanism of the SM. Therefore, via the mass of the heavy gauge boson $Z$ this mechanism can also gives minimal masses to all eight fermions. These minimal fermion masses originate naturally from basic physical energy considerations of the weak interaction. Therefore we obtain a nontrivial diagonal mass matrix as follows
\begin{equation}
\label{eq:15}
\tilde{\texttt{M}} = \mathrm{diag}\,[\tilde{m}_u,\tilde{m}_u,\tilde{m}_u,m_\nu,\tilde{m}_d,\tilde{m}_d,
\tilde{m}_d,\tilde{m}_e] \propto \upsilon,
\end{equation}
which indicates its origin of the Higgs vacuum. Thus the present ideas explain why the intuitive assumption of Yukawa coupling, which is merely made ad hoc in the SM about the fermion mass origin, may have indeed a deeper physical meaning. The minimal fermion masses consist just of the self-energies in the $Z_\mu$ gauge field, to which all fermions couple owing to their weak charges.

\begin{table}
	\caption{Fermion masses ($m_e=511$\,keV/$c^2$) induced by the self-energy of the electrostatic field} 
	\label{tab:2}
	\begin{tabular}{ccccc}
		\hline
		& up quark & neutrino & down quark & electron \\
		\hline
		Symbol & $\bar{m}_u$ &  $\bar{m}_\nu$ & $\bar{m}_d$ & $m_e$ \\
		Formula & $(Q_u/e)^2 m_e$ & $(Q_\nu/e)^2 m_e $ & $(Q_d/e)^2 m_e$ & Eq.\,(\ref{eq:2}) \\
		Value & 0.44 & 0.00 & 0.11 & 1.00 \\
		\hline
	\end{tabular}
\end{table}

Since the neutrino does not take part in the electromagnetic and strong interactions, it will keep its initial tiny mass unchanged, whereas the other fermions do not. As discussed in the beginning of this paper, the electron mass $m_e$ stems, owing to its electric charge, from its Coulomb self-energy, but is not further affected by the strong interaction. Adding to it the small mass $\tilde{m}_e$ does not make a difference. Our above reasoning was built on exploiting that the masses $m_e$ and $m_\nu$ are known from measurements. In contrast, the up and down quarks are dominated by the strong interaction and thus their constituent masses will also be largely determined by the colour-charged but massless gluon fields \cite{wilczek2008}. However, the electrostatic part of their masses can be estimated by help of equations (\ref{eq:5}) and (\ref{eq:6}). Following the initial considerations about the electron mass, we obtain another instructive Table~\ref{tab:2}.

Similarly, we obtain a nontrivial diagonal electrostatic mass matrix as follows
\begin{equation}
\label{eq:16}
\bar{\texttt{M}} = \mathrm{diag}\,[\bar{m}_u,\bar{m}_u,\bar{m}_u, 0,\bar{m}_d,\bar{m}_d,
\bar{m}_d, m_e],
\end{equation}
Again, the resulting ``electrostatic masses" $\bar{m}_{u,d}$ for the quarks are about ten to hundred times smaller than the constituent masses $m_{u,d}$ as inferred from QCD in the SM. Since the quarks are confined and thus do not exist as free particles, the electrostatic parts of their masses cannot be directly measured, yet may be helpful to know as estimates for a good starting point, when considering lattice simulations within quantum chromodynamics. 

Let us return finally to the extended Dirac equation (\ref{eq:17}) with the gauge fields. As the weak contributions of the masses (\ref{eq:15}) are much smaller then the electrostatic ones (\ref{eq:16}), we can simplify the diagonal mass matrix, by considering this and by putting also the initially assumed arbitrary common mass $m$ equal to zero. Then we obtain with $m_{u,d} \gg \bar{m}_{u,d} + \tilde{m}_{u,d}$ as the final result the diagonal mass matrix
\begin{equation}
	\label{eq:20}
	\texttt{M} = \mathrm{diag}\,[m_u, m_u, m_u, m_\nu, m_d, m_d, m_d, m_e]. 
\end{equation}

In conclusion, both mass matrices being Kronecker products do of course still commute with the five Gamma matrices of the extended Dirac equation \cite{mana2024} without gauge fields. But when including them, the chiral $SU(2)$ and the hadronic $SU(4)$ symmetries both are broken by the mass term. Therefore, breaking of the unified $SU(2) \otimes SU(4)$ symmetry by the Higgs mechanism gives the fermions their different charges and specific masses. They originate physically from the major self-energy of the electrostatic field as well as the minor self-energy contributed by the $Z$-boson field in proportion to its Higgs vacuum entering via the Compton wavelength that involves the Higgs boson mass. In contrast, the masses of the composite hadrons, in particular of the proton and neutron, involve dominant contributions from the energy of the binding gluon fields, as the QCD lattices simulation have clearly shown \cite{wilczek2008}.

\section{Conclusions}
\label{sec:9}

In this letter, we have considered a new intuitive idea of how the elementary fermions might acquire their finite empirical masses. We obtained diagonal mass matrices as Kronecker products within the framework of the unified gauge-field model of Marsch and Narita \cite{mana2024}. The mass matrices still commute with the five Gamma matrices of the extended free Dirac equation without gauge fields. However, when including them the chiral $SU(2)$ and the hadronic $SU(4)$ symmetries both are broken by the mass term. Thus, the breaking of the initial unified $SU(2) \otimes SU(4)$ symmetry by the Higgs-like mechanism gives the fermions their different charges as well as specific masses.

In the SM the initial common mass $m$ is assumed to be zero, and then the Dirac spinor splits into two independent two-component Weyl spinors. But when the gauge fields are switched on, their self-energy gives inertia and thus mass to the fermions in their rest frame. The breaking of gauge symmetry yields the electromagnetic massless photon field $E_\mu$ and the weak boson field $Z_\mu$, which becomes very massive via the Higgs mechanism. It also induces inertia for all eight fermions, yet the resulting masses are rather small owing to the very small Compton wavelength of the $Z$ boson. The neutrino and electron can acquire masses in this way, which yet differ by six orders of magnitude. The hadronic charge of the leptons is zero, and thus they decouple entirely from QCD. It is responsible by confinement through the gluons for the mass of the various resulting composite fermions, in particular for the proton mass.

The masses of the light fermions are thus argued to originate physically from the major self-energy of the electrostatic field as well as from the minor self-energy of the $Z$-boson field, which is proportional to the Higgs vacuum that determines the $Z$-boson mass. It is clear, however, that the masses of heavy composite hadrons, in particular of the proton and neutron, involve dominant contributions from the energy of the binding gluon fields, as the QCD lattice simulations have clearly shown \cite{wilczek2008}.

In conclusion, the extended Dirac equation (\ref{eq:17}) contains a physically well motivated mass term. It remedies the shortcoming of the SM that assumes massless fermions at the outset, whereas the empirical reality indicates that they are all massive. Therefore, the neutrino cannot be a Majorana particle, as it has often been suggested in the literature. This notion is in obvious contradiction to the observed neutrino oscillation \cite{mcdonald2004,kajita2006}, implying clearly finite masses. Chiral symmetry is broken in our theory, yet the parity remains intact. 

Finally, we like to mention the masses of the heavy gauge bosons involved in the above covariant derivative ({\ref{eq:24}) and related matrix ({\ref{eq:25}). In the reference of the particle data group \cite{pdg2022} we find in units of MeV/$c^2$ the values: $M_Z=91.2$ and $M_W=80.4$. For the ``leptoquark" boson $V$ we \cite{mana2024} obtain $M_V=35.4$. For the sum of these masses we find the following surprising results: $M_V + M_Z=126.6$, which equals within less than a one-percent margin the measured mass of the Higgs boson, $M_H=125.3$. Also, $M_W + M_Z=171.6$, which again equals within less than a one-percent margin the measured mass of the top quark, $M_T=172.7$. Whether this is just a fortuitous coincidence or indicates a physical connection has to remain open.


\appendix

\section{Appendix: CPT symmetry}
\label{sec:10}

We recall that the charge conjugation of the standard Dirac equation \cite{schwartz2014} is defined in the Weyl basis as 
$\mathcal{C}= \mathrm{i} \gamma_\mathrm{y} \mathrm{C}$, with the complex conjugation operator C, which transforms $\mathrm{i}$ into $-\mathrm{i}$. $\mathcal{C}$ is its own inverse, i.e. $\mathcal{C}^{-1}=\mathcal{C}$. Here we generalize this operation to have the form $\mathcal{C}= \mathrm{i} \gamma_\mathrm{y} \otimes {\sf{1}_8} \mathrm{C}$. When operating on a hermitian matrix, $\mathcal{C}$ is defined to also include a transposition of the $8 \times 8$ matrix involved. Thus $\mathcal{C} \texttt{C}_\mu \mathcal{C}^{-1} =\texttt{C}_\mu^\dagger=\texttt{C}_\mu$, as the gauge fields are real. However, for the standard Dirac matrices one obtains $\mathcal{C}\gamma^\mu \mathcal{C}^{-1}=-\gamma^\mu$. Then operation with the charge conjugation on (\ref{eq:17}) yields
\begin{equation}
	\label{eq:19} 
	\gamma^\mu \otimes (\mathrm{i} \partial_\mu {\sf{1}_8} - \texttt{S}_\mu(x)) \Psi_\mathcal{C}(x) 
	= {\sf{1}_4}\otimes\texttt{M} \,\Psi_\mathcal{C}(x).
\end{equation}
The charge-conjugated spinor is given by $\Psi_\mathcal{C}(x) = (\mathrm{i}\gamma_\mathrm{y} \otimes {\sf{1}_8}) \Psi^*(x)$. It obeys the same Dirac equation as (\ref{eq:17}), yet the sign of the charges $g_2$ and $g_4$, as well as of $e$ in (\ref{eq:18}) is now negative, like it should be for antiparticles.

Chiral symmetry is broken, yet parity is still obeyed by the extended Dirac equation, and so is time reversal. The standard Dirac equation remains invariant by application of the parity operator 
$\mathcal{P}=\gamma_0 \mathrm{P}$, with P sending $\mathbf{x}$ to $\mathbf{-x}$, $\mathcal{P}^{-1}=\mathcal{P}$. Similarly, it remains invariant under time reversal, $\mathcal{T}=\gamma_\mathrm{x} 
\gamma_\mathrm{z} \mathrm{T}$, with T changing $t$ into $-t$, and $\mathcal{T}^{-1}=-\mathcal{T}$. Note however, that $\mathcal{P}\gamma^\mu \mathcal{P}^{-1}=\gamma_\mu$, and $\mathcal{T}\gamma^\mu 
\mathcal{T}^{-1}=\gamma_\mu$ as well. Since the gauge fields transform like four-vectors, one finds that $\mathcal{P} Z_\mu(t,\mathbf{x}) \mathcal{P}^{-1}=Z^\mu(t,-\mathbf{x})$, and $\mathcal{T} Z_\mu(t,\mathbf{x}) \mathcal{T}^{-1}= Z^\mu(-t,\mathbf{x})$. Finally, we have $\mathcal{P} \partial_\mu \mathcal{P}^{-1}= \partial^\mu$, and $\mathcal{T} \partial_\mu \mathcal{T}^{-1}= \partial^\mu$. The related transformed spinors are $\Psi_\mathcal{P}(x) = (\gamma_0 \otimes {\sf{1}_8}) \Psi(t,-\mathbf{x})$, and 
$\Psi_\mathcal{T}(x)=(\gamma_\mathrm{x}\gamma_\mathrm{z} \otimes {\sf{1}_8}) \Psi(-t,\mathbf{x})$. Using these results one can show that the validity of the important $\mathcal{P}\mathcal{C}\mathcal{T}$ theorem is ensured.

\acknowledgments

%


\end{document}